\documentclass[preprint]{aastex631}

\begin{document}

\title{The Fast and Furious in JWST high-$z$ galaxies}

\correspondingauthor{Maurice H.P.M. van Putten}
\email{mvp@sejong.ac.kr}
\author[0000-0002-0786-7307]{Maurice H.P.M. van Putten}
\affiliation{Physics and Astronomy, Sejong University\\
209 Neungdong-ro, Seoul, South Korea }

\affiliation{INAF-OAS Bologna, via P. Gobetti, 101, I-40129 Bologna, Italy}

\begin{abstract}
Recent JWST surveys reveal a striking abundance of massive galaxies at cosmic dawn, earlier than predicted by $\Lambda$CDM.
The implied speed-up in galaxy formation by gravitational collapse is reminiscent of short-period galaxy dynamics described by the baryonic Tully-Fisher relation. This may originate in weak gravitation tracking the de Sitter scale of acceleration $a_{dS}=cH$, where $c$ is the velocity of light and $H(z)\propto \left(1+z\right)^{3/2}$ is the Hubble parameter with redshift $z$. With no free parameters, this produces a speed-up in early galaxy formation by an order of magnitude with essentially no change in initial galaxy mass function. It predicts a deceleration parameter $q_0=1-\left( 2\pi/GAa_{dS}\right)^2 =  -0.98\pm 0.5$, where $G$ is Newton's constant and 
$A=(47\pm6)M_\odot$\,(km/s)$^{-4}$ is the baryonic Tully-Fisher coefficient (McGaugh 2012). At $3\sigma$ significance, it identifies dynamical dark energy alleviating $H_0$-tension when combined with independent $q_0$ estimates in the Local Distance Ladder.  
Conclusive determination of $q_0=d\log(\theta(z)H(z))/dz\left|_{z=0}\right.$ is expected from BAO angle $\theta(z)$ observations by the recently launched {\em Euclid} mission.
\end{abstract}



\section{Introduction}

Recent JWST surveys \citep{eis23,aus23} 
reveal an abundance of massive galaxies in the Early Universe. 
Extending previous HST high-redshift galaxy surveys \citep{coe13,oes16} 
This early galaxy formation faster than expected poses 
a radically new challenge to $\Lambda$CDM 
\citep{eis05,boy23,mel23}
in addition to the $H_0$-tension problem between the Local Distance Ladder
and the Planck $\Lambda$CDM analysis of the {\em Cosmic Microwave Background} (CMB) seen in the recent history of cosmological expansion 
\citep{agh20,rie22}.

Galaxies form by gravitational collapse of relic density perturbations at the time of the surface of last scattering 
\citep{egg62,gun72,got75,got77}, 
marking large-scale structure at the angular scale of {\em Baryon Acoustic Oscillations} (BAO) in the CMB 
\citep{boy23,mel23,pad23,gup23}. 
Early galaxy formation by fast collapse is reminiscent of short-period orbital motion in spiral galaxies described by the Tully-Fisher luminosity-velocity relation 
\citep{tul77}.
Alternative interpretations of the latter by dark matter 
\cite{des17,wec18} 
and non-Newtonian physics
\citep{mil83} 
appear viable but breaking the degeneracy between the two is challenging 
\citep{fam12,mcg12}.

A common origin to early galaxy formation and the baryonic Tully-Fisher relation points to non-Newtonian dynamics in weak gravitation
{\em even when applied to dark and baryonic matter combined}. This appears inevitable in the face of the essentially matter-dominated evolution in the early Universe satisfying $\Omega_M\simeq 1$, defined by the ratio of total matter density over closure density $\rho_c=3H^2/8\pi G$ with Hubble parameter $H$ and Newton's constant $G$.

Generally, galaxy dynamics is mostly in weak gravitation below the de Sitter scale of acceleration $a_{dS}=cH$, defined by the Hubble parameter $H$ and the velocity of light $c$. Weak gravitation beyond $\Lambda$CDM may be tracking $a_{dS}$ with observational consequences by redshift dependence
\citep{van17}.

First, it predicts a non-smooth transition to non-Newtonian acceleration. Circular orbits in spiral galaxies can be described by $a_N/\alpha$ of expected Newtonian acceleration $a_N$ to the observed centripetal acceleration $\alpha$ as a function of $\zeta = a_N/a_{dS}$. These have a $C^0$-transition at $\zeta = 1$, observed in a $6\,\sigma$ gap between data from the {\em Spitzer Photometry and Accurate Rotation Curves} (SPARC) and $\Lambda$CDM galaxy models 
\citep{van18}. 

This $C^0$-transition can be attributed to a change in the binding energy in the gravitational field of centripetal acceleration $\alpha$ by the equivalence principle. In the Newtonian limit, the inertia of this binding energy equals a particle's rest mass. 
This is cut short when accelerations drop below $a_{dS}$ ($\zeta$ drops below unity), as the Rindler horizon at $\xi = c^2/\alpha$ drops beyond the Hubble radius $R_H=c/H$ \citep{van17}.  
This cut gives a $C^0$-transition at $\zeta=1$ with corresponding radius
\begin{eqnarray}
r_t=\sqrt{R_HR_G} \simeq 4.7\,\mbox{kpc}\,M_{11}^{1/2}\propto (1+z)^{-3/4}
\label{EQN_rt}
\end{eqnarray}
in galaxies of mass $M=10^{11}M_{11}M_\odot$ with gravitational radius $R_G=GM/c^2$. Accordingly, high-$z$ galaxies are increasingly non-Newtonian.  

The asymptotic regime $\zeta \ll1$ of dynamics in disks of spiral galaxies satisfies the baryonic Tully-Fisher relation of total mass $M_b$ in gas and stars and rotation velocities $V_c$, satisfying \citep{mcg12}
\begin{eqnarray}
M_b = A V^4_c
\label{EQN_BTFR}
\end{eqnarray} 
with $A\simeq \left(47\pm 6\right)M_\odot$\,(km/s)$^{-4}$. 
In the absence of dark matter, (\ref{EQN_BTFR}) 
is described by centripetal accelerations 
$\alpha=V_c^2/r$ at radius $r$, 
that can be modeled in terms of a logarithmic potential. 
Specifically, (\ref{EQN_BTFR}) is equivalent to Milgrom's law $\alpha=\sqrt{a_0a_N}$ for short-period orbital motion beyond the Newtonian orbits at $a_N=GM_b/r^2$, 
parameterized by $a_0=1/GA \simeq \left( 1.6\pm 0.2\right) 10^{-10}/\mbox{s}^2$ 
\citep{mil83,mcg12}.
However, redshift dependence in $A(z)$ is inconclusive. High-resolution observations are limited to galaxies at essentially zero redshift 
\citep{lel19} 
and recent observers out to moderate redshifts 
\citep{gen17} 
measure rotation curves limited to moderate radii, intermediate between the asymptotic acceleration $1/r$ in Milgrom's law and Newtonian acceleration $1/r^2$ 
\citep{van18}.

\begin{figure}
\centerline{
\includegraphics[scale=0.4]{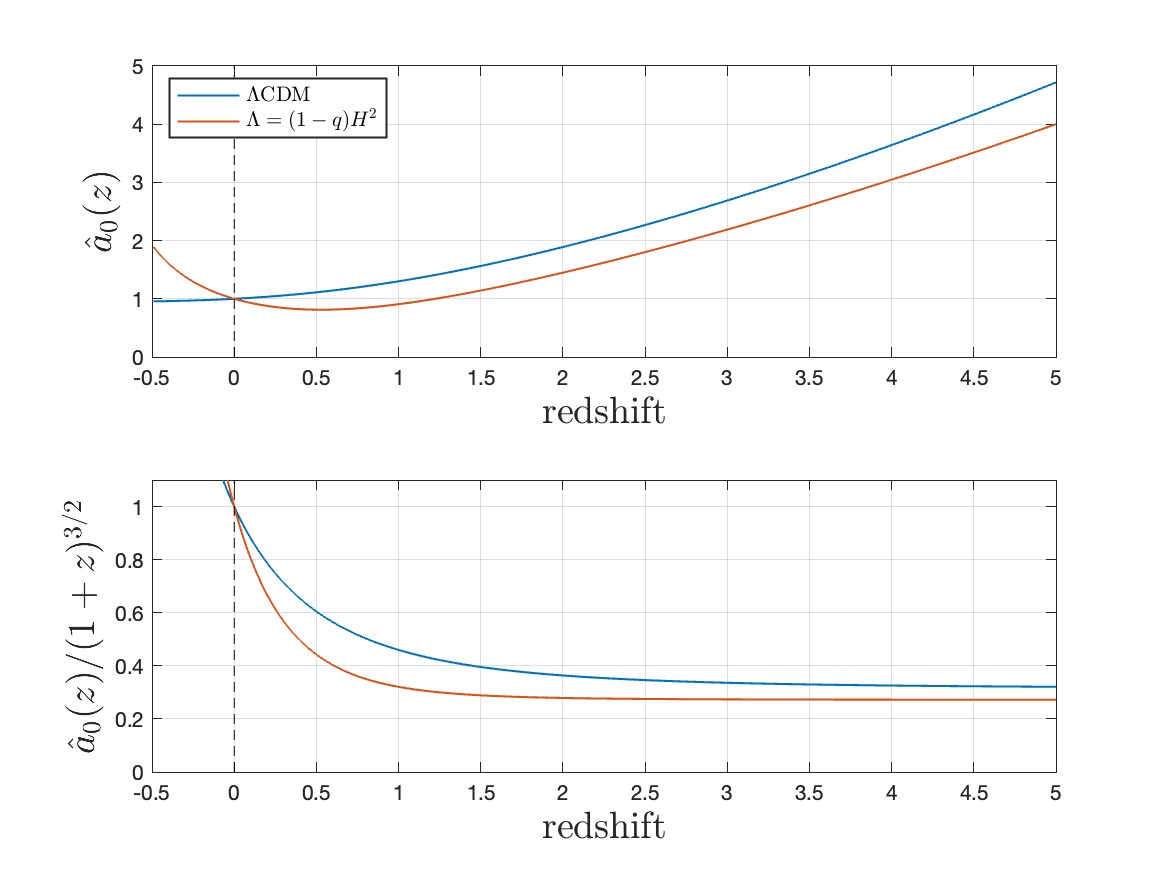}
}
\caption{Cosmological evolution of $\hat{a}_0(z)=a_0(z)/a_{0,0}$, normalized to the present-day value $a_{0,0}=a_0(0)$, shown for a background with constant dark energy ($\Lambda$CDM) and a dynamical dark energy $\Lambda=(1-q)H^2$ (the trace of the Schouten tensor) with no $H_0$-tension. 
While $\hat{a}_0(z)$ increases appreciably in late-time $\Lambda$CDM, it varies only slightly in the second.
}
\label{fig1}
\end{figure}

Second, weak gravitation $r\gg r_t$ tracking $a_{dS}$ in (\ref{EQN_rt}) predicts redshift dependence in the Milgrom parameter (Fig. \ref{fig1})
\begin{eqnarray}
a_0(z) = \frac{\sqrt{1-q(z)}}{2\pi} a_{dS}(z) 
\propto \left(1+z\right)^{3/2}~~\left(z\gg1\right),
\label{EQN_a0}
\end{eqnarray}
where $q(z)=-1+(1+z)H^{-1}H^\prime(z)$ is the deceleration parameter in a three-flat Friedmann universe.

Here, we show that (\ref{EQN_a0}) produces fast galaxy formation at cosmic dawn, explaining the JWST observations with no free parameters. Matter-dominated cosmological evolution covers an extended epoch up down to intermediate redshifts before the onset of the presently observed accelerated expansion 
\citep{rie98,per99,agh20}. 
This epoch includes primeval galaxy and structure formation since the surface of last scattering ($z\simeq 1100$) when cosmological evolution is described by a Friedmann scale factor $a(t)\propto t^{2/3}$, $a=1/(1+z)$ as a function of cosmic time $t$. 

In \S2, we discuss fast gravitational collapse in weak gravitation defined by $a_{dS}$. The consequences for accelerated galaxy formation are quantified in \S3. In \S4, we summarize our findings with a new estimate of $q_0$ in tension with $\Lambda$CDM.

\section{Fast gravitational collapse}

Critical to galaxy formation is the time scale of gravitational collapse and the time scale of formation of the first stars and dynamical relaxation. 
Expansion in the early universe might encompass more time as a function of redshift \citep{gup23,mel23}. 
This can be modeled by a stretched time $\tau$. For instance, power law $a\propto \tau^n$ gives additional time over $\Lambda$CDM by a factor
\begin{eqnarray}
\frac{\tau(z)}{t(z)}=\left(1+z\right)^\frac{3n-2}{2n} = \sqrt{1+n},
\label{EQN_n}
\end{eqnarray}
where the right-hand side exemplifies linear expansion ($n=1$) 
\citep{mel23}.

For the ultra-high redshift JWST galaxies, $\sqrt{1+z}\sim  4$ appears to be sufficient to satisfy 
observational constraints on the initial formation rate of the first stars and galaxies 
\citep{mel23}. 
However, precision cosmology at low redshifts including the ages of the oldest stars in globular clusters are pose stringent constraints on alternative cosmological models. In particular, a cosmology $a(t)\propto t$, has a deceleration parameter $q_0=0$, which is securely ruled out by data of the Local Distance Ladder and Planck $\Lambda$CDM analysis of the CMB 
\citep[e.g.][]{abc21}.

Nevertheless, (\ref{EQN_n}) for $n=1$ sets a scale for the required speed-up in galaxy formation soon after the Big Bang to accommodate the recent JWST observations. A gain similar to (\ref{EQN_n}) 
obtains equivalently from dynamical time scales that are sufficiently short, i.e., gravitational collapse times faster than expected from $\Lambda$CDM. 

Gravitational collapse takes place on a free-fall time scale. Representative analytic expressions are obtained in the two-body problem of radial motion of a test particle to a central mass $M$ at initial separation $R_0$ with zero initial momentum. For Newton's force law $1/r^2$ and $1/r$ in the baryonic Tully-Fisher relation (\ref{EQN_BTFR}), the respective free-fall time scales are $t_{ff}^{(2)} = {\pi}{2^{-3/2}} \left( {R_0} / {a_{N,0}} \right)^{1/2} \propto R_0^{3/2}$ and
$\tau_{ff}^{(2)} = {\sqrt{\pi}}{2^{-1/4}} \left( {R_0}/{\sqrt{a_0a_{N,0}}}\right)^{1/2} \propto R_0$.
Distinct from Newton's theory, the second is not scale-free by coupling to $a_{dS}$, leading to scaling with initial separation different from Newton's theory which points to relatively short free-fall time scales. While attributed to reduced inertia, this outcome is equivalently modeled by constant (Newtonian) inertia in a logarithmic potential. 
The ratio of the free-fall times hereby satisfies 
\begin{eqnarray}
\tau_{ff}^{(2)} = \frac{2^{5/4}}{\sqrt{\pi}}\left(\frac{a_{N,0}}{a_0}\right)^{1/4} t_{ff}^{(2)},
\label{EQN_tr}
\end{eqnarray}
explicitly showing faster gravitational collapse in the regime of weak gravitation, when $a_{N,0}\ll a_{dS}$.

\section{Accelerated galaxy formation}

\begin{figure}
\centerline{
\includegraphics[scale=0.4]{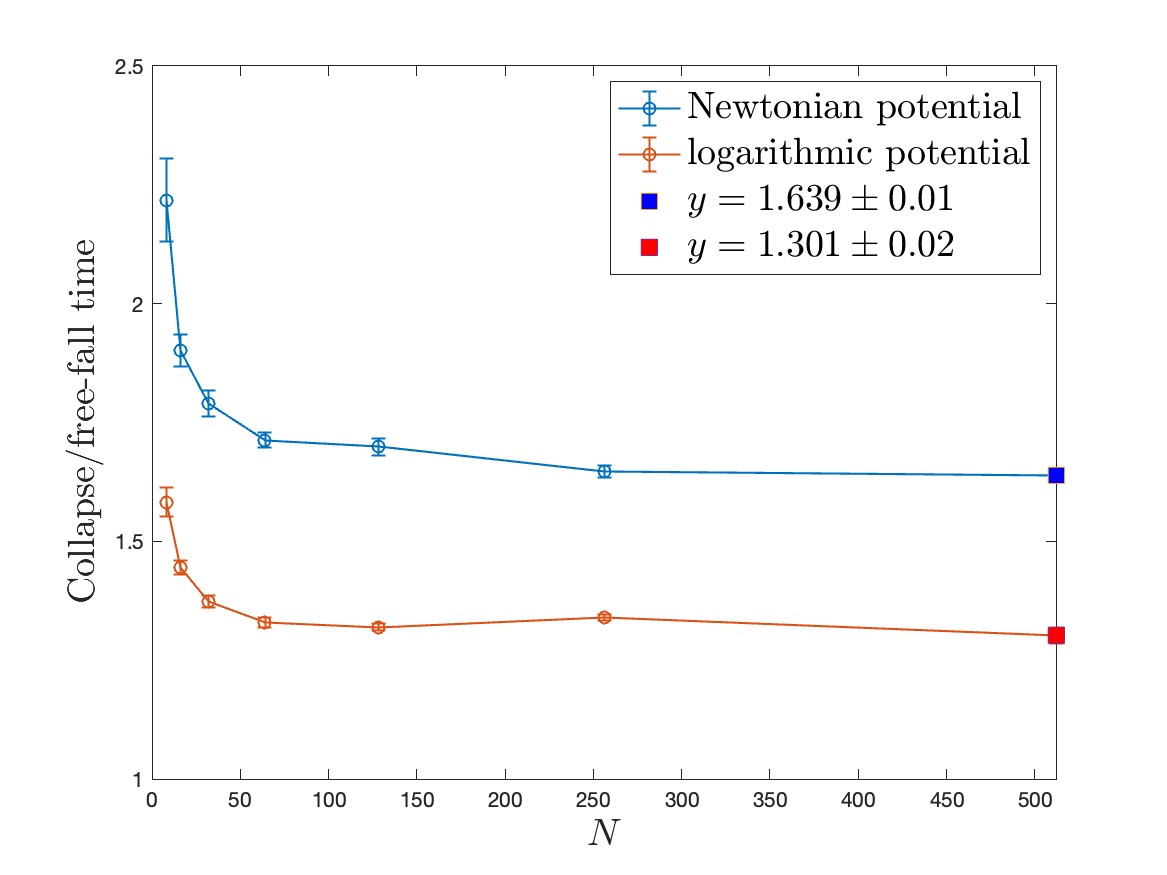}
}
\caption{The gravitational collapse times of an initially cold system shown for Newtonian and logarithmic potentials by direct
numerical simulations with Gaussian initial distributions of $N$ particles. The logarithmic potential models reduced inertia in the asymptotic regime of weak gravitation tracking $a_{dS}$. 
Both collapse times scale with $N^{-1/2}$, here shown by normalization to the Newtonian free-fall time scale $t_{ff}$. 
Generally, collapse times increase by the Hubble flow, more so for the Newtonian than for the logarithmic potential.
}
\label{fig2}
\end{figure}

We reinterpret (\ref{EQN_n}) in terms of fast gravitational collapse by (\ref{EQN_tr}). 
By Gauss' law, the Newtonian result in (\ref{EQN_tr}) permits a direct generalization $a_{N,0}=GNm/R_0^2$ for an (initially spherically symmetric) system of $N$ particles of mass $m$, 
the same does not apply to the logarithmic potential. 
Hence, the time of gravitational collapse (defined by the time to first bounce prior to virialization) is evaluated numerically. 
In the large $N$-limit, $N$-body simulations of initially cold clusters show (Fig. \ref{fig2}) 
the relations $t_c = 1.64\,t_{ff}$ and $\tau_c=1.30\,t_{ff}$
for the Newtonian and, respectively, logarithmic potential in (\ref{EQN_a0}), 
the later used to model reduced inertia in weak gravity below $a_{dS}$. 
Here, $t_{ff}= {\pi \left< R^{3/2}_0\right>}/{2\sqrt{2 N}}$ is the free-fall time-scale of an $N$-body cluster with mean $\left< R^{3/2}_0\right>$ of the initial particle radial positions $R_0$. Importantly, $t_{ff}$ expresses scaling with $N^{-1/2}$, inferred from the exact two-body free-fall time $t_{ff}^{(2)} = ({\pi}/{2}) \left( {R_0}/{2a_{N,0}}\right)^{1/2}$  $a_{N,0}=GM/R_0^2$ initial Newtonian acceleration of a test particle by gravitational attraction at initial separation $R_0$ to a mass $M$.
This scaling with $N^{-1/2}$ for both potentials 
can be attributed to the tight correlation with virtualization time, the latter dominated by diffusion. 

On a cosmological background, the $N$-body scaling relations (Fig. \ref{fig2}) and (\ref{EQN_tr}) show a speed-up in gravitational collapse
\begin{eqnarray}
B=\frac{t_c}{\tau_c} = 0.94 \left(\frac{a_0}{a_{N,0}}\right)^{1/4}= 
1.62 \,\zeta^{-1/4},
\label{EQN_c}
\end{eqnarray}
where $\zeta = a_N/a_{dS}$ \citep{van18}. 
Here, $q=1/2$ holds on the matter-dominated era at cosmic dawn. 
Specifically, (\ref{EQN_c}) applies to the scale $l$ of progenitor mass taken from a perturbation in background closure density $\rho_c$ feeding the formation of a $M=10^9M_9M_\odot$ galaxy, i.e.: 
\begin{eqnarray}
l\simeq \left(\frac{2M}{\rho_{c,0}\sqrt{\eta \Omega_{M,0}}}\right)^{1/3}\frac{1}{1+z}\simeq \left(\frac{l_0}{1+z}\right) M_9^{1/3},
\end{eqnarray} 
where $l_0 \simeq 150\,$\mbox{kpc}.
By $a_N=GM/l^2\simeq (4\pi/3)\rho_cl=(1/2)H^2l$,
$\zeta = a_N/a_{dS}=(1/2)\beta$, $\beta = Hl$. 
Speed-up in gravitational collapse (\ref{EQN_c}) hereby satisfies
\begin{eqnarray}
B \simeq 25 M_9^{-\frac{1}{12}}(1+z)^{1/8},
\label{EQN_B}
\end{eqnarray}
where $\beta_0 = l_0H_0/c\simeq 3.6\times 10^{-5}$ denotes the local Hubble flow across $l_0$.

\section{Conclusions}

In weak gravitation coupled to $a_{dS}$, (\ref{EQN_B}) shows collapse times to be considerably shorter than the Newtonian time scales of $\Lambda$CDM. Here, this is parameterized with no free parameters by (\ref{EQN_a0}). This predicts weak gravitation to govern most of the gravitational interactions in galaxy formation early on identified by JWST. Speed-up (\ref{EQN_B}) in gravitational collapse is over an order of magnitude, sufficient to account for the JWST observations. 

With no free parameters, we derive (\ref{EQN_B}) from a natural unification of the baryonic Tully-Fisher relation in late-time cosmology and fast galaxy formation at cosmic dawn.
Crucially, (\ref{EQN_B}) is essentially achromatic given a remarkably small power law index 1/12 in dependency on galaxy mass, which leaves galaxy mass distributions of $\Lambda$CDM effectively unchanged. 

The origin of (\ref{EQN_B}) is found in accelerated dynamics beyond the $C^0$-transition (\ref{EQN_rt}) upon identifying 
inertia with binding energy in the gravitational field of the accelerating particle (by the equivalence principle). 
On a cosmological background with a finite Hubble radius $R_H$, 
weak gravitation hereby has a finite sensitivity to $a_{dS}$. 
Expressed by (\ref{EQN_a0}), it applies to the asymptotic regime $\alpha \ll a_{dS}$, when binding energy (whence inertia) is cut short by the Hubble horizon
\citep{van17}. 

For the first time, (\ref{EQN_a0}) is confronted at high redshift by JWST observations. The resulting speed-up in galaxy formation (\ref{EQN_B}) takes us closer to cosmic dawn, beyond what can be explained by $\Lambda$CDM galaxy models, in a unification with the baryonic Tully-Fisher relation (\ref{EQN_BTFR}). This confrontation gives clear evidence of galaxy dynamics tracking $a_{dS}$
\citep{van17}.
 
The large speed-up (\ref{EQN_B}) points to the existence of galaxies beyond those currently observed by JWST. These may be detected in upcoming JWST surveys or by ultra-high redshift gamma-ray bursts with the planned {\em Transient High-Energy Sky and Early Universe Surveyor} (THESEUS) mission 
\citep{ama18,ama21}.

In the more recent epoch of cosmic expansion, (\ref{EQN_a0}) expresses sensitivity to the deceleration parameter $q(z)$ as it drops from matter-dominated ($q\simeq 1/2$) to negative values, signifying accelerated expansion during the present dark energy-dominated epoch. Fundamental to the latter is the combination $(H,q)$, parameterized by the present-day values $(H_0,q_0)$ of the Hubble constant $H_0$ and deceleration constant $q_0$. Inverting (\ref{EQN_a0}) gives 
\begin{eqnarray}
q_0 = 1- \left(\frac{2\pi }{GA a_{dS}}\right)^2 = -0.98^{+0.60}_{-0.42}
\label{EQN_q0}
\end{eqnarray}
given the baryonic Tully-Fisher coefficient of ({\em 20}) in (\ref{EQN_BTFR}) and $H_0=73.3\,$km\,s$^{-1}$Mpc$^{-1}$. 

The estimate (\ref{EQN_q0}) is consistent with $q_0=-1.08\pm 0.29$ derived from the Local Distance Ladder \citep{oco19}. 
An even combination of these two independent measurements gives 
$q_0=-1.03\pm 0.17$ distinct from the Planck $\Lambda$CDM value $q_0\simeq 0.5275$ ({\em 5}). At $3\sigma$ significance, this evidences a dynamical dark energy alleviating $H_0$-tension 
\citep{van18,n1,oco19,van20}

A decisive result on $q_0$ - and hence the nature of dark energy - is expected from a survey of the recent expansion history of the Universe by the recently launched {\em Euclid} mission ({\em 35}). Specifically, {\em Euclid} planned measurement of the BAO angle $\theta(z)$ and $H(z)$ will provide the radically new measurement 
\begin{eqnarray}
q(z) = (1+z)\frac{d}{dz}\log \left(\theta(z)H(z)\right).
\label{EQN_qz}
\end{eqnarray}
The expected Euclid survey of ${\cal O}\left(10^9\right)$ galaxies out to a redshift of a few
is expected to rigorously distinguish between dynamical dark energy ($q_0\simeq -1$, $H^\prime(0)\simeq0$) from $\Lambda$CDM 
($q_0\simeq -0.5$, $H^\prime(0)\simeq H_0$) and to reveal the physical nature of low-energy quantum cosmology 
and the problem of stability of de Sitter space \citep{van21}. 

\section{Acknowledgments}
\begin{acknowledgments}
We thank M.A. Abchouyeh for stimulating discussions. 
\end{acknowledgments}



\end{document}